\documentclass[aps,twocolumn,showpacs,preprintnumbers,floatfix,amssymb]{revtex4} 

\usepackage{epsfig}
\usepackage{graphicx}

\newcommand{\rf}[4]{{#1} {\bf #2}, #3 (#4)}

\newcommand{\beq}{\begin{equation}}
\newcommand{\eeq}{\end{equation}}
\newcommand{\beqa}{\begin{eqnarray}}
\newcommand{\eeqa}{\end{eqnarray}}

\begin{document}

\preprint{
\hbox{ADP-03-143/T578}
}

\title{Topological Charge Evolution in the Markov-Chain of QCD}

\author{Derek B. Leinweber}
\email{dleinweb@physics.adelaide.edu.au}
\homepage{http://www.physics.adelaide.edu.au/theory/staff/leinweber/}

\author{Anthony G. Williams}
\email{anthony.williams@adelaide.edu.au}

\author{Jian-bo Zhang}
\email{jzhang@physics.adelaide.edu.au}

\affiliation{Department of Physics and
         Special Research Centre for the Subatomic Structure of Matter,
         University of Adelaide 5005, Australia}

\author{Frank X. Lee}
\email{fxlee@gwu.edu}

\affiliation{Center for Nuclear Studies, Department of Physics,
        The George Washington University, Washington, D.C. 20052 and \\
        Jefferson Lab, 12000 Jefferson Avenue,
        Newport News, VA 23606}

\begin{abstract}
The topological charge is studied on lattices of large physical volume
and fine lattice spacing.  
We illustrate how a parity transformation on the $SU(3)$
link-variables of lattice gauge configurations reverses the sign of
the topological charge and leaves the action invariant.  Random
applications of the parity transformation are proposed to traverse
from one topological charge sign to the other.  The transformation
provides an improved unbiased estimator of the ensemble average and is
essential in improving the ergodicity of the Markov chain process.
\end{abstract}


\maketitle


\section{Introduction}
\label{sec:intro}

The Markov chain process is the corner stone of modern
importance-sampling techniques for generating field configurations in
the numerical simulation of quantum field theories.  Central to the
process is ergodicity; the ability to move through configuration space
from any particular field configuration to any other field
configuration with finite probability.  Vacuum
expectation values of observable operators are estimated from the
average of field configurations selected with a probability given by
the exponentiation of the Euclidean action, $\exp(-S_E)$,
governing the quantum field theory.  

It is essential that these field configurations are selected based on
the action alone, unbiased by the field configuration used to initiate
the Markov chain.  In practice this is done by updating the field
configurations for thousands of sweeps through the lattice, monitoring
the autocorrelation of various observables, and selecting a new
representative field configuration only after significant evolution
through configuration space.

The topological charge of a gauge configuration in lattice QCD has
already been identified as a particularly difficult quantity to evolve
in the Markov chain process, displaying unusually long autocorrelation
times \cite{Alles98, dawson, longAutoDBW1,longAutoDBW2}.  Some gauge
actions such as the renormalization-group block-transform based DBW2
action \cite{dbw2-1,dbw2-2} are notorious for locking in the
topological charge at fine lattice spacings
\cite{longAutoDBW1,longAutoDBW2}.  Difficulties associated with the
Iwasaki gauge action \cite{Iwasaki:we,Iwasaki:1983ck} are presented
here.

In this study we address an aspect of the topological charge evolution
problem that will present itself for any gauge field action.  In
particular, we address a problem associated with the approach to the
infinite-volume continuum limit in numerical supercomputer simulations
of $SU(3)$ gauge theory and QCD in general.

As the physical volume, $V$, of the lattice increases, the average
value of the squared topological charge increases, in accord with
the topological susceptibility, given by
\beq
\chi = \frac{\langle Q^2 \rangle}{V} \, .
\eeq
In quenched QCD, $\chi$ is related to physical hadron masses via the
large $N_c$ Witten-Veneziano relation
\cite{Witten:1979vv,Veneziano:1979ec}
\beq
\chi = \frac{f_\pi^2}{2\, N_f} \left ( m_\eta^2 + m_{\eta'} - 2\,
m_K^2 \right ) \, .
\eeq
In full QCD, the quark mass dependence of the topological
susceptibility may be related to pseudoscalar meson properties via the
Gell-Mann--Oakes--Renner relation as
\cite{Crewther:1977ce,DiVecchia:1980ve,Leutwyler:1992yt,Hart:2001pj}
\beq
\chi = \frac{f_\pi^2 \, m_\pi^2}{2\, N_f} + {\cal O}(m_\pi^4) \, .
\eeq
On a sufficiently large volume lattice, the distribution of the
topological charge is expected to be Gaussian.  Regions of non-trivial
topological charge density are uncorrelated for sufficiently large
separations and a normal distribution will result.  The distribution
of Q for the Wilson gauge action has been found to be Gaussian to a
very good approximation for both quenched QCD
\cite{DelDebbio:2002xa,DelDebbio:2002gr} and full QCD where the Wilson
gauge action is complemented by the Wilson fermion action
\cite{Alles98} or the Wilson-clover action \cite{Hart:2004ij}.

\begin{figure}[t]
\begin{center}
{\includegraphics[angle=90,width=0.95\hsize]{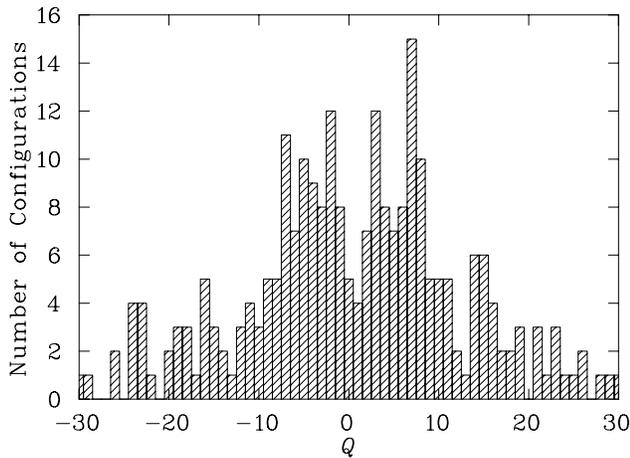}}


\caption{Histogram illustrating the number of configurations in our
  ensemble of 250 configurations having a particular topological
  charge, $Q$.  A suppression of $Q \sim 0$
  configurations is observed.
}
\label{Qhist} 
\end{center}
\end{figure}

\begin{figure}[t]
\begin{center}
{\includegraphics[angle=90,width=0.99\hsize]{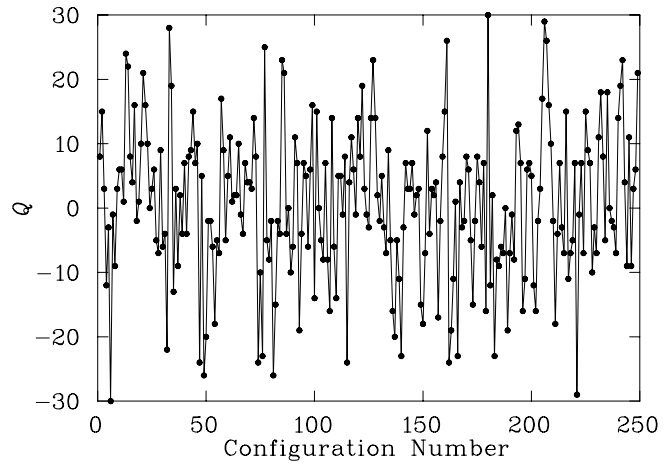}}
\caption{The time evolution of the topological charge plotted as a
  function of simulation time as represented by the configuration
  number.  1000 pseudo-heatbath sweeps are performed between each
  gauge field sample.}
\label{Qtime} 
\end{center}
\end{figure}

Alternate distributions of the topological charge can occur in a
finite volume if the correlation length of the topological charge
density for a particular lattice gauge action approaches the length of
the lattice dimensions.  Figure \ref{Qhist} displays a histogram of
the topological charge from an ensemble of 250 quenched Iwasaki
\cite{Iwasaki:we,Iwasaki:1983ck} gauge configurations at $\beta =
9.1674 $ on a $28^3 \times 44$ lattice, where the lattice spacing $a$
is $0.113$ fm.  The topological charge is calculated using the
highly-improved, three-loop ${\cal O}(a^4)$-improved lattice
field-strength tensor \cite{Bilson-Thompson:2002jk} on 10-sweep cooled
configurations obtained with a three-loop ${\cal O}(a^4)$-improved
action.  The gauge configurations represented in Fig.~\ref{Qhist} are
separated by 1000 pseudo-heatbath sweeps in an attempt to obtain
uncorrelated configurations.  Fig.~\ref{Qtime} illustrates the time
evolution of the topological charge plotted as a function of
simulation time, represented by the configuration number.  Acceptable
movement of the topological charge is indicated by the rapidity and
amplitude of the oscillations.  To the best of our knowledge, this is
the first time a double-peaked structure in the probability
distribution of the topological charge has been revealed for a
renormalization-group improved gauge action.  It would be interesting
to examine this distribution for the DBW2 action.

For any lattice action, the distribution of the topological charge will
broaden as the volume of the lattice increases, such that
\beq
\langle Q^2 \rangle = \chi\, V \, .
\label{broad}
\eeq
As the infinite-volume continuum limit is approached and correlation
lengths diverge, it will become increasingly difficult to evolve the
topological charge over the broad range demanded by Eq.~(\ref{broad}).
In particular, a symmetric distribution about $Q=0$ is required to
preserve the symmetries of QCD.

Recognizing the need for improvement in the evolution of the
topological charge in the Markov chain process, we present a simple
transformation in the following section, that may be applied to a
gauge-field configuration of quenched QCD or dynamical-fermion QCD
that changes the sign of $Q$ while leaving the action invariant.  The
transformation provides an improved unbiased estimator of the ensemble
average and is essential in improving the ergodicity of the Markov
chain process.

\section{Gauge-Link Transformation}
\label{sec:ParityT}

The index theorem \cite{as68} relates the topological charge $Q$ to the
chirality index of the Dirac operator $D$ on a continuum
4-torus as
\beq
 Q = {\rm index}(D) \equiv n_- - n_+  \, . 
\label{AS_theorem}
\eeq
Here $n_+$ and $n_-$ are the number of exact zero eigenmodes, $D\psi =
0$, with positive, $\gamma_5 \psi = + \psi$, (right-handed) and
negative, $\gamma_5 \psi = - \psi$, (left-handed) chiralities
respectively.  

This link between the topological charge and the chirality (or
helicity) of zero eigenmodes of the Dirac operator identifies parity
as the transformation for changing the sign of the topological charge
while leaving the action invariant.  Helicity transforms as a
pseudoscalar under rotations, whereas the action of QCD, designed to
conserve parity, transforms as a scalar under rotations.  For the
improper rotation of the parity transformation, the right-handed modes
will become the left-handed, and vice versa, changing the sign of $Q$.

In deriving the transformation of the links under parity, we begin
with the transformation of the gauge potential.  Under the parity
transformation
\beq
A_i(\vec{x},t) \to -A_i(-\vec{x},t) \, ,~~\mbox{and}~~
\partial_{i} \to -\partial_{i} \, , 
\eeq
for the spatial indices $i = 1, 2, 3$ and
\beq
A_4(\vec{x},t) \to A_4(-\vec{x},t)\, ,~~\mbox{and}~~
\partial_{4} \to \partial_{4} \, . 
\eeq
Here the $3\times 3$ colour matrix degrees of freedom of $A_\mu(x)$
are implicit.  The non-abelian field strength tensor $F_{\mu\nu}$ is
\beq
F_{\mu\nu} = \partial_{\mu}A_{\nu}
-\partial_{\nu}A_{\mu} + i\, g\, \left [A_{\mu}, A_{\nu}\right ] \, .
\eeq 
Under parity transformations
\beq
F_{ij} \to F_{ij} \, ,~~F_{4j} \to -F_{4j} \, ,~~\mbox{and}~~
F_{i4} \to -F_{i4} \, .
\label{parityFmunu}
\eeq 
Hence the action
\beq
S = \frac{1}{2} \int d^4x\, {\rm Tr}
\left ( F_{\mu\nu}(x)\, F_{\mu\nu}(x) \right ) \, ,
\label{action}
\eeq
is invariant under parity while the topological charge 
\beq
Q = \frac{g^2}{32\pi^2} \int d^4x\,
\epsilon_{\mu\nu\rho\sigma} \, 
{\rm Tr}
\left ( F_{\mu\nu}(x)\, F_{\rho\sigma}(x) \right) \, ,
\label{topQ}
\eeq
changes sign due to the presence of $\epsilon_{\mu\nu\rho\sigma}$
ensuring the presence of one and only one time component in the field
strength tensor product.

The gauge-field links are related to the gauge potential via
\beq
U_{\mu}(x) = \exp \left ( i\, g \int_0^a \,A_{\mu}(x+\lambda \, \hat{\mu})
\, d\lambda \right ) \, ,
\eeq
where $ x \equiv (\vec{x},t)$.  Under the parity transformation, the
spatial components, $i = 1, 2, 3$, of the links transform as
\beqa
{\mathcal{P}}\, U_{i}(x)\, {\mathcal{P}^\dagger} &=& \exp \left ( -i\,
g \int_0^a  \, A_{i}(-\widetilde{x}-\lambda \, \hat{\imath}) \, d\lambda
\right ) \, ,  \\
&=&  U_i^\dagger(-\widetilde{x}-a\hat{\imath}) \, , 
\label{ptUs}
\eeqa
where 
\beq
-\widetilde{x} \equiv (-\vec{x},t) \, .  
\eeq
Similarly, the time components transform as
\beqa
{\mathcal{P}}\, U_{t}(x)\, {\mathcal{P}^\dagger} &=& \exp \left ( i \,
g \int_0^a  \, A_{t}(-\widetilde{x}+\lambda\, \hat{t}) \, d\lambda
\right ) \, , \\
&=& U_{t}(-\widetilde{x}) \, .
\label{ptUt}
\eeqa
The exact nature of the parity transformation on the links follows
from the fact that parity is an exact symmetry on the hypercubic
lattice. 

It is interesting to examine the manner in which the lattice action
density and lattice field-strength tensor transform under parity.
Consider the product of links about an elementary plaquette located at
space-time point $x$, in the $\mu$ - $\nu$ plane
\beq
U_{\mu \nu}(x) \equiv U_{\mu}(x)\, U_{\nu}(x+\hat{\mu})\,
U^{\dagger}_{\mu}(x+\hat{\nu})\, U^\dagger_{\nu}(x) \, ,
\eeq
and its transformation under parity.  For spatially oriented
plaquettes, the untransformed plaquette $U_{i j}(x)$ originates from $x$,
and loops counter-clockwise in the positive $\hat{\imath}$, $\hat{\jmath}$
direction
\beq
U_{i}(x)\, U_{j}(x+\hat{\imath})\, U^{\dagger}_{i}(x+\hat{\jmath})\,
U^\dagger_{j}(x) \, .
\eeq
Under the parity transformation $U_{i j}(x)$ transforms to
\beq
U^{\dagger}_{i}(-\widetilde{x}-\hat{\imath})\,
U^{\dagger}_{j}(-\widetilde{x}-\hat{\imath}-\hat{\jmath})\, 
U_{i}(-\widetilde{x}-\hat{\jmath}-\hat{\imath})\,  U_{j}(-\widetilde{x}-\hat{\jmath}) \, ,
\label{sspt}
\eeq
which is the spatial plaquette originating from $-\widetilde{x}$,
looping in a counter-clockwise orientation again, this time in the
negative $\hat{\imath}$, $\hat{\jmath}$ direction.  The space-time oriented
plaquettes originating from $x$ and looping counter-clockwise in the
positive $\hat{\imath}$, $\hat{t}$ direction
\beq
U_{i}(x)\, U_{t}(x+\hat{\imath})\, U^{\dagger}_{i}(x+\hat{t})\,
U^\dagger_{t}(x) \, , 
\eeq
transform to
\beq
U^{\dagger}_{i}(-\widetilde{x}-\hat{\imath})\,
U_{t}(-\widetilde{x}-\hat{\imath})\,
U_{i}(-\widetilde{x}-\hat{\imath}+\hat{t})\, U^\dagger_{t}(-\widetilde{x}) 
\, ,
\label{stpt}
\eeq
which is the space-time oriented plaquette originating from
$-\widetilde{x}$ but this time looping clockwise in the negative
$\hat{\imath}$, positive $\hat{t}$ direction.

The connection between these link products and the lattice field
strength tensor is provided by
\begin{eqnarray}
g\, F_{\mu\nu}(x) & = & \frac{1}{8\, i} \biggl [ 
\left ( {\cal{O}}_{\mu\nu}(x)-{\cal{O}}^{\dagger}_{\mu\nu}(x)
\right ) \nonumber \\
&&\qquad - \frac{1}{3} {\rm Tr} \left ( {\cal{O}}_{\mu\nu}(x) -
{\cal{O}}^{\dagger}_{\mu\nu}(x) \right) \biggr ] \, ,
\label{lattFmunu}
\end{eqnarray}
where ${\cal{O}}_{\mu\nu}(x)$ is the sum of ${1\times{1}}$ link paths
oriented about $x$ in the $\mu$ - $\nu$ plane
\begin{eqnarray}
\lefteqn{ {\cal{O}}_{\mu\nu}(x) = U_{\mu}(x)\, U_{\nu}(x+\hat{\mu})\,
	     U^{\dagger}_{\mu}(x+\hat{\nu}) \, U^\dagger_{\nu}(x) }
             \nonumber \\ 
	     &+& U_{\nu}(x)\,
	     U^{\dagger}_{\mu}(x+\hat{\nu}-\hat{\mu})\, 
             U^{\dagger}_{\nu}(x-\hat{\mu})\, U_{\mu}(x-\hat{\mu})
	     \nonumber \\ 
	     &+& U^{\dagger}_{\mu}(x-\hat{\mu})\,
	     U^{\dagger}_{\nu}(x-\hat{\mu}-\hat{\nu})\,
	     U_{\mu}(x-\hat{\mu}-\hat{\nu})\, U_{\nu}(x-\hat{\nu})
	     \nonumber \\ 
	     &+& U^{\dagger}_{\nu}(x-\hat{\nu})\,
	     U_{\mu}(x-\hat{\nu})\, U_{\nu}(x+\hat{\mu}-\hat{\nu})\,
	     U^{\dagger}_{\mu}(x) \, .
\label{Omunu}
\end{eqnarray}
Taking the Hermitian conjugate of ${\cal{O}}_{\mu\nu}(x)$ changes the
orientation of the link products from counter-clockwise to clockwise.
Noting further that $F_{\mu\nu}(x)$ is odd under such transformations,
we see that the orientations of the link-product parity
transformations of Eqs.~(\ref{sspt}) and (\ref{stpt}) are precisely those
required to recover the continuum transformations of $F_{\mu\nu}(x)$
in Eq.~(\ref{parityFmunu}).  While the topological charge density
undergoes a parity transformation, the magnitude of the topological charge
remains invariant, as all lattice sites are summed over.  Similarly, the lattice action 
\beq
S = \beta \sum_{\scriptstyle x \atop \scriptstyle \mu < \nu}
\frac{1}{6}\, {\rm Tr} \left ( 2 - U_{\mu \nu}(x) - U_{\mu
  \nu}^\dagger(x) \right ) \, , 
\eeq
is even under the orientation transformation $U_{\mu \nu}(x) \to
U_{\mu \nu}^\dagger(x)$, and since all sites are summed over, the
lattice gauge action is invariant under the lattice parity
transformation.

\section{ Numerical Results }
\label{sec:results}

The parity transformations of the gauge links in Eqs.~(\ref{ptUs}) and
(\ref{ptUt}) have been coded in Fortran 90 \cite{webAddress} and
tested on a $12^3\times 24$ lattice with the tadpole-improved
L\"{u}scher-Weisz gauge action \cite{Luscher:1984xn}.
We use three methods to examine the properties of the parity
transformed configurations.

The 3-loop improved field strength tensor \cite{Bilson-Thompson:2002jk}
accompanied by the associated 3-loop improved cooling action are used
to provide a determination of the topological charge $Q$ and action
$S$ for a cooled gauge configuration and its parity transformed
partner.  The transformed action and topological charge magnitude
agree to machine precision.

The low-lying spectral flow of the hermitian Wilson-Dirac operator
$H_W$ = $\gamma_5 D_W$ is examined.  Under the parity transformation,
the low-lying eigenvalues of $H_W$ change sign such that the slope of
the spectral flow (proportional to the topological charge giving rise
to the zeromode) changes sign.  Since $H_W$ is of even dimension
(lattice volume $\times$ number of colors $\times$ number of Dirac
indices), $\det(H_W) = \det(\gamma_5) \times \det(D_W)$ is unaffected
by the parity transformation, as required to preserve parity in
dynamical fermion QCD simulations.

Finally, the topological charge is determined by counting the exact
zeromodes of the massless overlap Dirac operator.  Under a parity
transformation, the number of exact zeromodes with positive
(right-handed) and negative (left-handed) chiralities exchange as
expected.

\section{Discussion and Conclusions}
\label{sec:conclusions}

The manner in which the parity transformation is used in practice will
depend on the application at hand for the gauge fields.  Often, exact
parity can be enforced during the construction of correlation
functions by averaging opposite-sign nontrivial momenta of the lattice
Green's functions \cite{Leinweber:1990dv}.  In other applications,
exact parity can be enforced by doubling the number of gauge field
configurations, averaging each field configuration with its parity
transformed partner.  Finally, the transform can be applied in the
Markov chain process itself, applying the transform randomly with a
50\% probability prior to writing the gauge configuration to disk, or
equivalently, following the reading of the configuration to be
evolved, from disk.

Motivated by the lattice index theorem, we have illustrated how a
parity transformation on the $SU(3)$ link-variables of lattice gauge
configurations reverses the sign of topological charge while leaving
the action invariant.  The transformation provides an improved
unbiased estimator of the ensemble average and is essential in
improving the ergodicity of the Markov chain process.

We thank Philippe De Forcrand, Alistair Hart and Ettore Vicari for
beneficial correspondence.  Supercomputing support from the Australian
National Facility for Lattice Gauge Theory, The South Australian
Partnership for Advanced Computing (SAPAC) and the Australian
Partnership for Advanced Computing (APAC) is gratefully acknowledged.
This research is supported by the Australian Research Council.


\end{document}